\documentclass[11pt]{article}
\usepackage{latexsym}  
\usepackage{amssymb}
\usepackage{graphicx}
\usepackage{amsmath}

\begin{document}

\begin{center}

{\Large\bf Family Gauge Boson Mass Estimated from $K^+ \rightarrow \pi^+ \nu \bar{\nu}$ }

\vspace{4mm}
{\bf Yoshio Koide }

{\it Department of Physics, Osaka University, 
Toyonaka, Osaka 560-0043, Japan}

{\it E-mail address: koide@kuno-g.phys.sci.osaka-u.ac.jp}

\end{center}

\vspace{3mm}

\begin{abstract}
It is emphasized that a rare decay $K^+ \rightarrow \pi^+ \nu \bar{\nu}$
becomes promising in a future search for a new particle, 
because the theoretical treatment is well established and the  
value of the branching ratio $Br(K^+ \rightarrow \pi^+ 
\nu \bar{\nu})$ is sensitive to a search for a new particle 
with a TeV scale mass. As an example, according to a U(3)  family 
gauge boson model which predicts the lowest family gauge boson 
with a few TeV mass $M_{11}$, the branching ratio  
$Br(K^+ \rightarrow \pi^+ \nu \bar{\nu})$ is discussed.
If we can obtain, in future, a slightly lower value 
$Br^{obs} \sim 0.9 \times 10^{-10}$ compared with the present 
observed value $Br^{obs}=(1.7\pm 1.1)\times 10^{-10}$, 
we can conclude $M_{11} \sim$ a few TeV.  
\end{abstract}

PCAC numbers:  
  11.30.Hv, 
  12.60.-i, 
  14.70.Pw, 

\vspace{5mm}

\noindent{\large\bf 1 \ Introduction} 

In the current physics, search for a new particle is one of our big concern. 
Except for direct searches (e.g. $Z'$ search at the LHC), it is usually 
tried to investigate a deviation between an observed value and its 
standard model (SM) prediction.  
However, in the harmonic events, it is not so easy to evaluate 
the QCD effects exactly. 
Leptonic processes will rather be important for such new 
particle searches. 
In this paper, we notice a rare decay $K^+ \rightarrow \pi^+ \nu \bar{\nu}$
as a promising way for such a new particle search. 

The present observed value \cite{PDG14} and a SM prediction 
\cite{Isidori_NPB05} are as follows:
$$
Br^{obs} \equiv Br(K^+ \rightarrow \pi^+ \nu \bar{\nu})_{obs} = 
(1.7 \pm 1.1) \times 10^{-10} ,
\eqno(1.1)
$$
$$
Br^{SM} \equiv Br(K^+ \rightarrow \pi^+ \nu \bar{\nu})_{SM} = 
(0.80 \pm 0.11) \times 10^{-10} .
\eqno(1.2)
$$
Of course, the observed value (1.1) is consistent with the SM value 
(1.2), so that there is no room of a new particle. 
However, the observed value (1.1) has a large error at present.
As far as the center values are concerned, we see a 
sizable deviation from each other. 
The predicted value (1.2) is reliable because the decay is a   
leptonic one, so that the theoretical treatment is well established. 
We would like especially to notice that, different from most 
rare decay searches in which we know only upper limits of the 
branching ratios, we know that the rare decay 
$K^+ \rightarrow \pi^+ \nu \bar{\nu}$ has really been observed 
although $Br^{obs}$ has a large error at present. 
We may expect that the error in the observed value would be reduced 
in the near future. 
The study would be hopeful as the nearest way of a new particle 
search.

In this paper,  we discuss a contribution of a family 
gauge boson (FGB) to the branching ratio 
$Br(K^+ \rightarrow \pi^+ \nu \bar{\nu})$ on the basis of a specific 
FGB model \cite{Koide_PLB14} which gives a possibility of a TeV scale 
mass of the lowest FGB (a brief review will be given later).   
We will conclude that if the observed value becomes near  
the SM prediction (1.2), i.e. if we could have  
$Br^{obs} \sim 0.9 \times 10^{-10}$ in future, we will be able 
to conclude that a mass  $M_{11}$ of the lowest FGB $A_1^1$ 
is an order of a few TeV.  
Of course, on the other hand, if the observed value becomes far from the 
SM value (1.2), i.e. $Br^{obs} >1.0 \times 10^{-10}$, 
the result will rule out the specific scenario of
FGBs given in Ref.\cite{Koide_PLB14}, and it will confirm the 
conventional FGB model. 
In either case, the experimental search for the rare decay 
$K^+ \rightarrow \pi^+ \nu \bar{\nu}$ will play an 
important role in investigating the FGB models. 

Meanwhile, we would like to talk about why we notice FGBs.
In the standard model (SM), the Cabibbo-Kobayashi-Maskawa (CKM) 
mixing matrix \cite{CKM} $V_{CKM}=U_u^\dagger U_d$ is, of course, observable 
quantity, while the up- and down-quark mixing matrices $U_u$ and
$U_d$ are not separately observable quantities.
When a family gauge symmetry exists, $U_u$ and $U_d$ can be individually
observable. 
We usually consider that 
masses  $M_{ij}$ of the family gauge bosons (FGBs) $A_i^{\ j}$ 
are larger than an order of $10^4$ TeV, because of the observed 
values of $P^0$-$\bar{P}^0$ mixings ($P=K, D, B, B_s$).
Therefore, even if FGBs exist, it will be hard to observe their effects 
by means of terrestrial experiments.  

Even though FGBs are different from a conventional model, we know 
a FGB model which has been proposed by Sumino \cite{Sumino_PLB09}
and which can give a considerably low mass of the lowest FGB.  
Sumino has introduced FGBs in order to solve a problem in a charged  
lepton mass relation \cite{K-mass} 
$K  \equiv ({m_e + m_\mu +m_\tau})/{\left(\sqrt{m_e}+ \sqrt{m_\tau} 
+\sqrt{m_\tau}\right)^2}  = {2}/{3}$: the mass relation   
is well satisfied by the pole masses [i.e. $K^{pole}=(2/3) \times
(0.999989 \pm 0.000014)$], while it is not so satisfactory for the
running masses [i.e. $K(\mu)=(2/3) \times (1.00189 \pm 0.00002)$ 
at $\mu = m_Z$].
The deviation is due to a factor $\log (m_{ei}^2/\mu^2)$ in the 
QED radiative correction.  
Sumino has considered that FGBs really exist, and the factor 
$\log (m_{ei}^2/\mu^2)$ is canceled by a factor 
$\log (M_{ii}^2/\mu^2)$ in the FGB contribution.  
In order that the cancellation works correctly as we hope,  
the FGB masses $M_{ii}$ have to be proportional 
to the charged lepton masses $m_{ei}$:
$$
M_{ij} = k \left( m_{ei}^n +m_{ej}^n \right)
\ \ \ (n=\pm 1, \pm 2, \cdots ), 
\eqno(1.3)
$$
and the family gauge coupling constant $g_F$ has to satisfy
a relation
$$
\left( \frac{g_F}{\sqrt2} \right)^2 =\frac{2}{n} e^2 =
\frac{4}{n} \left( \frac{g_w}{\sqrt2} \right)^2
\sin^2 \theta_w .
\eqno(1.4)
$$
(Although in the original Sumino model \cite{Sumino_PLB09}, 
$n$ has been taken as $n=1$, we can, in general, take 
$n= \pm 1, \pm 2, \cdots$.)
In the Sumino model, FGB mass matrix is diagonal in the
family basis in which the charged lepton mass matrix is diagonal.
Therefore, family number violation  does not occur in the 
lepton sector (at the tree level).
In contrast to the lepton sector, in the quark sector, 
the up- and down-quark mass matrices $M_u$ and $M_d$ are, in general,
not diagonal, so that family number violation appears at tree level
via quark mixing matrices $U_u$ and $U_d$.  (See Eq.(1.6) later.)
Therefore, FGB contribution to $P^0$-$\bar{P}^0$ mixing 
($P=K, D, B, B_s$) are allowed only through the quark mixings 
$U_u$ and $U_d$. 
This is a reason that the FGB contribution in the Sumino model 
is considerably suppressed compared with the conventional FGB model.
However, even in the Sumino FGB model, we cannot obtain a visible 
low mass with a TeV scale.  

Recently, a new FGB model \cite{Koide_PLB14} with a TeV scale mass of 
the lowest FGB has been proposed, without conflicting with the
observed $P^0$-$\bar{P}^0$ mixings, on the basis of the Sumino 
U(3) FGB model.  
The point of the success in the new model  \cite{Koide_PLB14}
compared with the original Sumino model \cite{Sumino_PLB09}
is as follow: 
when we define the family number in the lepton sector as 
$(e_1, e_2, e_3)=(e^-, \mu^-, \tau^-)$, 
the family number in the quark sector
is defined as $(d_1, d_2, d_3)=(d, s, b)$ in the Sumino model 
\cite{Sumino_PLB09} as the same as the conventional one, 
while it is defined as 
$$
(d_1, d_2, d_3)= (b, d, s) \ \ \ [{\rm or}\ 
 (d_1, d_2, d_3)= (b, s, d) ] ,
 \eqno(1.5)
$$
in the new model \cite{Koide_PLB14}.  
In addition to the original Sumino model, when 
the family number assignment is revised from the conventional
one to the new one (1.5), 
we can obtain a lower mass $M_{11}$ without conflicting with 
the observed $K^0$-$\bar{K}^0$ mixing.  
However, although a value $M_{11} \sim 1$ TeV has been reported 
in Ref.\cite{Koide_PLB14},  
the value should be not taken rigidly, because the estimates 
of $P^0$-$\bar{P}^0$ mixing are still dependent on QCD 
corrections and on many unfixed parameter values.  
We need a confirmation from another phenomenon.  

In conclusion, in this paper, according to the revised model 
\cite{Koide_PLB14} of the original Sumino model, we take the 
following interactions of quarks and leptons with FGBs:
$$
{\cal H}_{fam} = \frac{g_F}{\sqrt{2}} \left[ \sum_{\ell= \nu, e} 
\left( \bar{\ell}_L^i \gamma_\mu \ell_{Lj} - \bar{\ell}_{Rj} \gamma_\mu \ell_R^i
\right) +  \sum_{q=u,d} (U_q^{*})_{i k}
(U_q)_{j l} (\bar{q}_{k} \gamma_\mu  q_{l}) \right] (A_i^{\ j})^\mu ,
\eqno(1.6)
$$
together with the new family number assignment (1.5).
The first term (leptonic part) in Eq.(1.6) takes somewhat 
an unfamiliar form.
This is due to the following reason:
in the Sumino model, the minus sign for the cancellation
has been provided by the U(3) assignment 
$(e_L, e_R) \sim ({\bf 3}, {\bf 3}^*)$ for
the left-handed and right-handed charged leptons $e_L$ and $e_R$. 
As a result, we have unwelcome situation:   
(i) The model cannot be anomaly free.
(ii) Effective current-current
interactions with $\Delta N_{fam}=2$ ($N_{fam}$ is a family 
number) appear inevitably.   
In order to avoid problem (i), we tacitly assume an existence of 
heavy leptons in the lepton sector.
On the other hand, the problem (ii) causes a fatal damage to the 
$P^0$-$\bar{P}^0$ mixing problem.  
Therefore, only for quark sector, we restore the Sumino's 
assignment to the normal assignment 
$(q_L, q_R) \sim ({\bf 3}, {\bf 3})$
of U(3), because we do not need such the Sumino cancellation 
\cite{Sumino_PLB09} in the quark sector. 
Furthermore, in the present paper, according to the modified model 
\cite{Koide_PLB14}, we discuss the case $n=2$ which leads to  
a FGB mass $M_{11} \sim 1$ TeV in Ref.\cite{Koide_PLB14}.

\vspace{5mm}

\noindent{\large\bf 2 \ Estimate of $Br(K^+ \rightarrow \pi^+ \nu \bar{\nu})$ }

The value  ${M}_{11} \sim 1$ TeV in Ref.\cite{Koide_PLB14} has been 
estimated from a difference between the observed value of 
$K^0$-$\bar{K}^0$ mixing and its SM value. 
Of course, since the present observe value of $K^0$-$\bar{K}^0$ mixing 
is consistent with the predicted value (with a large error), so that 
the value $M_{11} \sim 1$ TeV should be regarded as a lower value 
in an optimistic estimate.  
On the other hand, the SM value is based on somewhat ambiguous input values 
(for example, QCD effects and so on).
In order to confirm this estimate, in this section, we estimate 
$Br(K^+ \rightarrow \pi^+ \nu \bar{\nu})$ which includes 
FGB contributions in addition to the SM contribution. 
In this estimate, we will use the value (1.2) as the SM contribution.
The estimated result will be compared with a value $Br^{obs}$ 
in a future observation, 
and thereby a possible value $M_{11}$ will be speculated.  

We assume that the difference between the values (1.1) and (1.2)
is due to contributions from FGBs, although the value (1.1) 
have large errors and it seems that the value is consistent 
with the SM prediction (1.2) at present. 
However, we consider that the error value in (1.1) will be 
improved in the near future.
Then, the value $M_{11}$ from $K^+ \rightarrow \pi^+ \nu \bar{\nu}$ 
will become rather reliable than that from the observed 
$K^0$-$\bar{K}^0$ mixing.

\begin{figure}[h]
\begin{center}
\begin{picture}(400,150)(0,-20)   
 \includegraphics[height=.12\textheight]{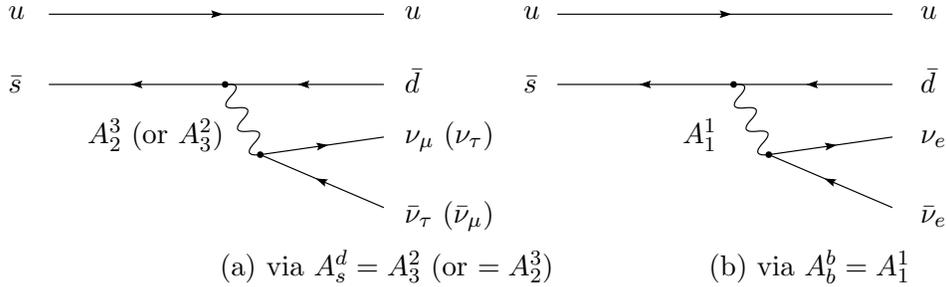}
 \hspace{20mm}
  \includegraphics[height=.12\textheight]{fig1.eps}
 \put(-335,72){$u$}
  \put(-335,44){$\bar{s}$}
 \put(-305,25){$A_2^3$ (or $A^2_3$)}
\put(-185,72){$u$}
\put(-185,44){$\bar{d}$}
\put(-185,25){$\nu_\mu$ ($\nu_\tau$)}
\put(-185,-5){$\bar{\nu}_\tau$ ($\bar{\nu}_\mu$)}
\put(-255, -25){(a) via $A_s^d=A_3^2$ (or $=A_2^3$)}
 \put(-140,72){$u$}
  \put(-140,44){$\bar{s}$}
 \put(-80,25){$A_1^1$}
\put(10,72){$u$}
\put(10,44){$\bar{d}$}
\put(10,25){$\nu_e$}
\put(10,-5){$\bar{\nu}_e$}
\put(-70, -25){(b) via $A_b^b=A_1^1$}

\end{picture}  
  \caption{
Contributions of FGBs in the rare decay $K^+ \rightarrow \pi^+ \nu \bar{\nu}$ 
}
\label{K-decay}
\end{center}
\end{figure}

The neutrinos $\nu \bar{\nu}$ in the SM prediction mean  
$\nu_e \bar{\nu}_e+\nu_\mu \bar{\nu}_\mu+\nu_\tau \bar{\nu}_\tau$
because it is caused by exchange of $Z$.
On the other hand, in the FGB contributions, there are two 
diagrams as seen in Fig.~1. 
A decay given in Fig.1 (a) is induced via $A_s^d$ (i.e. $A_2^3$ or $A_3^2$),
so that the final state $\nu \bar{\nu}$ means $\nu_\mu \bar{\nu}_e$. 
Another one [Fig.1 (b)] is induced via $A_1^1$, so that 
the final state $\nu \bar{\nu}$ means $\nu_e \bar{\nu}_e$. 
Those decay amplitudes  ${\cal M}_a$ and ${\cal M}_b$ are proportional 
to factors $1/M_{23}^2$ and $|(U_d)_{bd}| |(U_d)_{bs}|/M_{11}^2$, respectively.
Under the approximation $U_u \sim {\bf 1}$ and  $U_d \simeq V_{CKM}$, 
we obtain $|(U_d)_{bd}| |(U_d)_{bs}| = 3.59 \times 10^{-4}$.
On the other hand, we have FGB mass ratios 
$$
\frac{M_{23}^2}{M_{11}^2} = \frac{1}{2} \left( \frac{m_\tau^2}{
m_e^2} + \frac{m_\mu^2}{m_e^2} \right) 
= 6.05 \times 10^6 , 
\eqno(2.1)
$$
from Eq.(1.4) with $n=2$, 
so that we obtain the ratio of the matrix element ${\cal M}_b$
(Fig.1 (b)) to ${\cal M}_a$ (Fig.1 (a)) as
$|{\cal M}_b /{\cal M}_a| = 2.17  \times 10^3$.
Therefore, we find that the contribution from Fig.1 (a) is 
negligible compared with that from Fig.1 (b).
Hereafter we neglect the contribution from the diagram (a).

Let us denote the observed branching ratio $Br^{obs}\equiv
Br(K^+ \rightarrow \pi^+ \nu \bar{\nu})_{obs} $ as follows:
$$
Br^{obs} = k \left( |{\cal M}(\nu_e \bar{\nu_e})) |^2 +
|{\cal M}(\nu_\mu \bar{\nu_\mu} ) |^2 +
|{\cal M}(\nu_\tau \bar{\nu_\tau } )  |^2  \right) ,
\eqno(2.2)
$$
where ${\cal M} (\nu_i \bar{\nu_i})$ denotes matrix elements of 
$K^+ \rightarrow \pi^0 \nu_i \bar{\nu_i}$, and $k$ is a constant 
which includes phase volume for three body decay in the limit of
massless neutrinos.  
In the SM, since 
$$
{\cal M}(\nu_e \bar{\nu_e} )  = {\cal M}(\nu_\mu \bar{\nu_\mu} ) 
= {\cal M}(\nu_\tau \bar{\nu_\tau} ) \equiv  {\cal M}_{SM} ,
\eqno(2.3)
$$
the predicted branching ratio in SM $Br^{SM}$ can be expressed as
$$
Br^{SM} = 3 k |{\cal M}_{SM}|^2 .
\eqno(2.4)
$$
On the other hand, when we denote a matrix element corresponding to
Fig.1 (b) as ${\cal M}_{FGB} \equiv {\cal M}(\nu_e \bar{\nu_e} )_{FGB}$,
the observed branching ratio $Br^{obs}$ is given by
$$
Br^{obs} = k \left( ( |{\cal M}_{SM} + {\cal M}_{FGB} |^2 
+ 2 |{\cal M}_{SM}|^2 \right) 
= \frac{1}{3}  Br^{SM} \left(\left| 1+\frac{  {\cal M}_{FGB} }{{\cal M}_{SM} }
\right|^2 +2\right) .
\eqno(2.5)
$$
In order to estimate 
$$
(R_{11})^2 \equiv \left( \frac{  {\cal M}_{FGB} }{{\cal M}_{SM} } \right)^2
= 3 \frac{Br^{FGB}}{Br^{SM}} ,
\eqno(2.6)
$$
we define a parameter
$$
\varepsilon \equiv \frac{Br^{SM}(K^+ \rightarrow \pi^+ \nu \bar{\nu})}{
Br(K^+ \rightarrow \pi^0 \nu e^+)} .
\eqno(2.7)
$$
We use the following approximate relation
$$
\frac{
Br^{FGB}(K^+ \rightarrow \pi^+ \nu_e \bar{\nu}_e) }{
Br(K^+ \rightarrow \pi^0 \nu e^+) } = 
\frac{ |V_{td}|^2 |V_{ts}|^2 (g_F^2/8M_{11}^2)^2 f_+}{
\frac{1}{2} |V_{us}|^2 (g_W^2/8 M_W^2)^2 f_0} ,
\eqno(2.8)
$$
where $f_+ \equiv  f(m_{\pi^+}/m_{K^+})$, $f_0 \equiv f(m_{\pi^0}/m_{K^+})$,
and $f(x)$ is a phase space function 
$f(x) = 1 -8 x^2 + 8x^6 - x^8 -12 x^4 \log x^2$. 
Here, we have neglected the lepton masses. 
We have also neglected form factor effects in 
$K^+ \rightarrow \pi^+ \nu \bar{\nu}$ and 
$K^+ \rightarrow \pi^0 \nu_e e^+$.
We have assumed that both form factors are approximately similar, 
so that the effect is canceled in our estimate of the ratio.   
Therefore, we obtain
$$
R_{11} = \sqrt{\frac{6 }{\varepsilon }\,  \frac{f_+}{f_-}} \,  \xi 
\frac{g_F^2/8M_{11}^2}{ g_W^2/8 M_W^2} . 
\eqno(2.9)
$$
where $\xi \equiv  |V_{td}| |V_{ts}|/|V_{us}|= 1.59 \times 10^{-3}$ 
\cite{PDG14}. $f_+/f_0= 0.964$ and 
$g_W^2/8 M_W^2=1/2 v_H^2 =8.25$ TeV$^{-2}$ (we have sued $v_H=246.22$ GeV).
In the present paper, we will use the value of $\varepsilon$
$$
\varepsilon = \frac{(0.80 \pm 0.11) \times 10^{-10}}{
(5.07 \pm 0.04) \times 10^{-2}} = (1.58 \pm 0.22) \times 10^{-9} .
\eqno(2.10)
$$

Thus, by using Eqs.(2.5) - (2.10), we can estimate a value of 
$M_{11}$ for given value of $Br^{obs}$. 
For example, for $Br^{SM} = 8.0 \times 10^{-10}$ and 
$Br^{obs}= (1.7 \pm 1.1) \times 10^{-10}$ gives
$$
M_{11} = 0.54^{+\, \infty}_{-0.13} \, {\rm TeV} .   .
\eqno(2.11)
$$
(The expression (2.11) is somewhat misleading. 
The value $M_{11}=0.54$ TeV means a value of $M_{11}$ 
estimated by using the center value $Br^{obs} =1.7 \times 10^{-10}$.
On the other hand, the value $M_{11}=(0.54 -0.13) =0.41$ TeV means 
a value estimated by using the upper value 
$Br^{obs} = (1.7+1.1) \times 10^{-10} =2.8 \times 10^{-10}$.  
The value $M_{11} = \infty$ means nothing but that 
there is no room for a new physics at present.  
In the expression ${\ }^{+\sigma}_{-\sigma}$, the values 
mean neither the theoretical error nor the experimental error. 
Nevertheless, for the convenience, we will use this expression 
hereafter.)   
If the observed value of $Br^{obs}$ becomes slightly improved in future, 
for example, 
$Br^{obs}= (1.7 \pm 0.8) \times 10^{-10}$, we will obtain 
a finite value of $M_{11} = (0.54^{+0.82}_{-0.11} ) \times 10^{-10}$.

\begin{figure}[h]
\begin{center}
\begin{picture}(300,300)(0,0)   
 \includegraphics[height=.4\textheight]{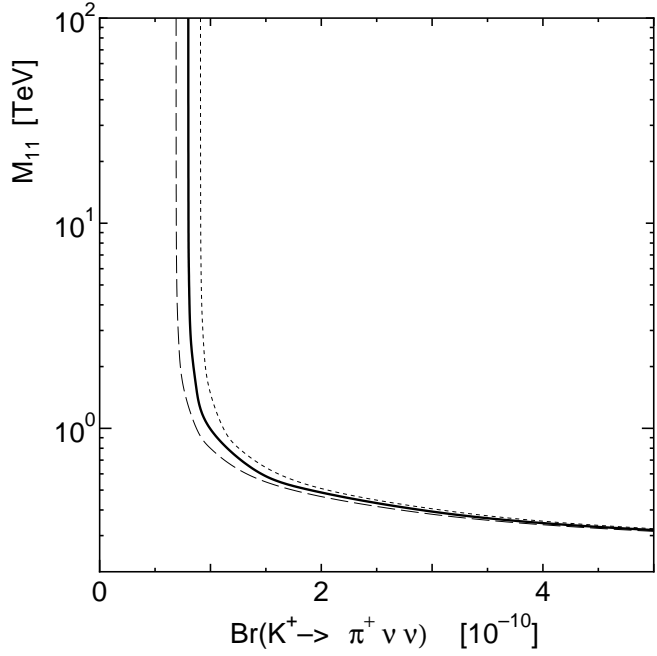}
  
\end{picture}  
  \caption{
Mass $M_{11}$ versus branching ratio $Br(K^+ \rightarrow \pi^+ \nu \bar{\nu})$. 
Solid, dashed and dotted curves denote cases of 
$Br^{SM}(K^+ \rightarrow \pi^+ \nu \bar{\nu})=0.80 \times 10^{-10}$,  
$(0.80+ 0.11) \times 10^{-10}$ and $(0.80 - 0.11) \times 10^{-10}$, 
respectively.  
If we take the present observed value $Br^{obs}=(1.7 \pm 1.1) \times 10^{-10}$
we obtain $M_{11}=0.54^{+\, \infty}_{-0.13}$ TeV for $Br^{SM}= 0.80
\times 10^{-10}$.
}
\label{br}
\end{center}
\end{figure}

However, an estimate of $M_{11}$ for larger than a TeV mass 
should be careful, because it is highly dependent on the value
of $Br^{SM}$. 
In order to understand this situation, we show the behavior of 
$M_{11}$ versus $Br(K^+ \rightarrow \pi^+ \nu \bar{\nu})$ 
in Fig.2.
As seen in Fig.2, it is impossible to estimate an exact value of 
$M_{11}$ under the SM prediction (1.2) with the present error.  

Nevertheless, we can conclude some important results from Fig.2:
(i) A lower limit of $Br^{obs}$ (not a upper limit) gives 
an upper limit of $M_{11}$.
A mass value $M_{11}$ smaller than 1 TeV has already 
ruled out by dilepton searches at the LHC \cite{ATRAS_PRD14}.
This suggests that if $A_1^{\ 1}$ really exits, an  observed 
value of $Br^{obs}$ in the near future must be smaller than 
$1 \times 10^{-10}$. 
In other words, if $Br^{obs} > 1.0 \times 10^{-10}$ is 
established, the specific FGB model discussed in this section 
will be ruled out.   
(ii) In order to estimate the value of $M_{11}$, more improvements 
of the errors are required not only for the observed value $Br^{obs}$
but also the SM estimation $Br^{SM}$.

So far we are taken $n=2$ in the FGB mass relation (1.3), because 
the case can give the mass $M_{11}$ with a few TeV without 
conflicting with the observed $P^0$-$\bar{P}^0$ mixings 
$=K, D, B, B_s$ \cite{Koide_PLB14}.
For reference, here, we would like to comment on 
the simplest case $n=1$. 
Since the case gives a mass ratio $(M_{23}/M_{11})^2 =1.84\times 10^3$
instead of the value (2.1), we cannot neglect the contribution given in 
Fig.1 (a) compared with Fig.1 (b). 
Then, we have to modify $Br^{obs}$ given in Eq.(2.2) as 
$Br(K^+ \rightarrow \pi^+ \nu \bar{\nu})_{obs} $ as follows:
$$
Br^{obs} = k \left( |{\cal M}(\nu_e \bar{\nu_e})) |^2 +
|{\cal M}(\nu_\mu \bar{\nu_\mu} ) |^2 +
|{\cal M}(\nu_\tau \bar{\nu}_\tau)  |^2 +
|{\cal M}(\nu_\mu \bar{\nu}_\tau)|^2 
 \right) ,
\eqno(2.12)
$$
(Case of $A_2^{\ 3}$ ($=A_d^s$)). 
Therefore, Eq.(2.5) is modified into
$$
Br^{obs} 
= \frac{1}{3}  Br^{SM} \left(\left| 1+R_{11}\right|^2 +2
+|R_{23}|^2 \right) ,
\eqno(2.13)
$$
where $R_{11}$ is defined by Eq.(2.6) and $R_{23}$ is
given by
$$
|R_{23}|^2 = 3 \frac{Br^{FGB}(\nu_\mu \bar{\nu}_\tau) }{Br^{SM}}= 
\frac{3}{\varepsilon}  \frac{(g_F^2/8 M_{23}^2)^2}{
\frac{1}{2}|V_{us}|^2 (g_W^2/8 M_W^2)^2 } .
\eqno(2.14) 
$$
The relation between $Br^{obs}$ and $M_{11}$ is also similar to Fig.2, 
but the numerical result is slightly changed.
For example, for the branching ratio $Br^{obs}=(1.7 \pm 1.1) \times 10^{-10}$
gives 
$$
M_{23}= 0.65^{+\, \infty}_{-0.14} \ {\rm TeV} .
\eqno(2.15)
$$
(For a hypothetical value  $Br^{obs}=(1.7 \pm 0.8) \times 10^{-10}$, 
we obtain $M_{11}=0.65^{+ 0.80}_{- 0.12}$ TeV. )
Although the lowest FGB mass is almost the same as the case 
of $n=2$ as far as we see the decay $K^+ \rightarrow \pi^+ \nu \bar{\nu}$, 
we should note that the strong constraint in the case $n=1$ still comes 
from the observed $K^0$-$\bar{K}^0$ mixing. 
The result (2.15) suggests only that the constraint from 
$K^+ \rightarrow \pi^+ \nu \bar{\nu}$ is not essential for the case $n=1$.

\vspace{5mm}

\noindent{\large\bf3 \ Concluding remarks }

According to a specific FGB model in which the FGB masses are related 
to the charged lepton masses as shown in Eq.(1.3), 
the family gauge coupling constant $g_F$ is not free as shown in Eq.(1.4), 
and the family number assignment in quark sector 
is twisted as shown in Eq.(1.5).  
We have estimated the lowest FGB mass $M_{11}$ from the deviation
between the observed value (1.1) and the SM estimate value (1.2). 
We cannot regrettably obtain meaningful value of $M_{11}$ 
because of the present large errors in $Br^{obs}$ and $Br^{SM}$.
However, we expect improvement of those values in future, 
because the rare decay $K^+ \rightarrow \pi^+ \nu \bar{\nu}$ has
really been detected differently from other lepton flavor 
changing rare decays. 
On the other hand, the estimate of $Br^{SM}$ is reliable because the decay is 
leptonic one, so that the theoretical treatment is well established.

However, the present estimate was done. basing on non-conventional model.  
As a reference, let us estimate a FGB mass on the basis of 
 conventional FGB model. 
In the conventional model in which the family number is assigned 
as $(d_1, d_2, d_3)= (d, s, b)$, a FGB which can dominantly contribute
to the decay $K^+ \rightarrow \pi^+ \nu \bar{\nu}$ is $A_1^2$ 
as ${\cal M}(K^+ \rightarrow \pi^+ \nu_e \bar{\nu}_\mu)$.
Therefore, there is no interference with the final states in 
the SM estimate: 
$$
Br^{obs} (K^+ \rightarrow \pi^+ \nu \bar{\nu}) = 
Br^{SM} (\nu_i \bar{\nu}_i) + Br^{FGB}(\nu_e \bar{\nu}_\mu) .
\eqno(3.1)
$$
Therefore, we can estimate in Sec.2, similarly,
$$
\frac{Br^{FGB}}{Br^{SB}} = \frac{1}{\varepsilon} \frac{(g_F^2/8 M_{12}^2)^2
f_+}{ \frac{1}{2} |V_{us}|^2 (g_W^2 /8 M_W^2)^2 f_0}, 
\eqno(3.2)
$$
where $\varepsilon$ is defined by Eq.(2.7).
Present values (1.1) and (1.2) lead to 
$$
\tilde{M}_{12} \equiv  \frac{M_{12}}{g_F/\sqrt2}  = 66.6^{+\,\infty}_{-1.1} 
\ {\rm TeV} .
\eqno(3.3)
$$
(Do not confuse $A_1^{\ 2}$ in the conventional model with 
$A_1^{\ 2}$ in the model \cite{Koide_PLB14} discussed in Sec.2. 
The FGB $A_1^{\ 2}$ in the conventional model means $A_d^{\ s}$ 
in terms of names of down-quarks, while $A_1^{\ 2}$ in Sec.2 
means $A_b^{\ d}$ [or  $A_b^{\ s}$].) 
Note that the gauge coupling constant $g_F$ is unknown parameter in 
the conventional FGB model, so that we can determine only the value of 
$\tilde{M}_{12}$ directly (not $M_{12}$).

This value (3.3) is comparatively low, so that the value is within 
reach of $\mu$-$e$ conversion experiments in preparation \cite{mu-e_conv}.  
However, such a low value (3.3) in the conventional model will conflict 
with the observed $K^0$-$\bar{K}^0$ mixing even we adopt 
the Sumino RGB model \cite{Sumino_PLB09}.
Only the way in which we can obtain FGBs with lower masses is still to adopt 
the twisted assignment of quark family numbers \cite{Koide_PLB14} in addition 
to the Sumino FGB model. 

In conclusion, the improvement of the values in  $Br^{obs}$ and 
$Br^{SM}$ seems to be very important to the estimate of FGB masses. 
If the observed value becomes close to 
the SM prediction (1.2), e.g. if we can have  
$Br^{obs} \sim 0.9 \times 10^{-10}$ in future, we will be able 
to conclude that a mass of the lowest FGB, $M_{11}$, is an 
order of a few TeV. 
If the observed value becomes inversely far from the 
SM value (1.2), e.g. $Br^{obs} > 1.0 \times 10^{-10}$, 
the result will rule out the specific scenario given 
in this paper, because such a value of $Br^{obs}$ leads to too low 
value of $M_{11}$, so that the case is ruled out by the data 
$pp \rightarrow e^+ e^- +X$ at the LHC \cite{ATRAS_PRD14}.
Such a value $Br^{obs}$ will, rather, confirm the conventional 
FGB model. 
In either case, the improvement of the observation of 
$Br(K^+ \rightarrow \pi^+ \nu \bar{\nu})$ will play an 
important role in the FGB searches.
We are eager for an improved observation of 
$K^+ \rightarrow \pi^+ \nu \bar{\nu}$.

\vspace{5mm}
%

%

\end{document}